# Topology-Level Reactivity in Distributed Reactive Programs

## Reactive Acquaintance Management using Flocks

Sam Van den Vonder[a], Thierry Renaux[a], and Wolfgang De Meuter[a]

a    Software Languages Lab, Vrije Universiteit Brussel, Belgium

**Abstract**    Reactive programming is a popular paradigm to program event-driven applications, and it is often proposed as a paradigm to write distributed applications. One such type of application is *prosumer* applications, which are distributed applications that both produce and consume many events.

We analyse the problems that occur when using a reactive programming language or framework to implement prosumer applications. We find that the assumption of an open network, which means prosumers of various types spontaneously join and leave the network, can cause a lot of code complexity or run-time inefficiency. At the basis of these issues lies *acquaintance management*: the ability to discover prosumers as they join and leave the network, and correctly maintaining this state throughout the reactive program. Most existing reactive programming languages and frameworks have limited support for managing acquaintances, resulting in accidental complexity of the code or inefficient computations.

In this paper we present acquaintance management for reactive programs. First, we design an *acquaintance discovery* mechanism to create a *flock* that automatically discovers prosumers on the network. An important aspect of flocks is their integration with reactive programs, such that a reactive program can correctly and efficiently maintain its state. To this end we design an *acquaintance maintenance* mechanism: a new type of operator for functional reactive programming languages that we call "deploy-*". The deploy-* operator enables correct and efficient reactions to time-varying collections of discovered prosumers.

The proposed mechanisms are implemented in a reactive programming language called Stella, which serves as a linguistic vehicle to demonstrate the ideas of our approach. Our implementation of acquaintance management results in computationally efficient and idiomatic reactive code.

We evaluate our approach quantitatively via benchmarks that show that our implementation is efficient: computations will efficiently update whenever a new prosumer is discovered, or a connected prosumer is dropped. To evaluate the distributed capabilities of our prototype implementation, we implement a use-case that simulates the bike-sharing infrastructure of Brussels, and we run it on a Raspberry Pi cluster computer.

We consider our work to be an important step to use functional reactive programming to build distributed systems for open networks, in other words, distributed reactive programs that involve many prosumer devices and sensors that spontaneously join and leave the network.

# The Art, Science, and Engineering of Programming

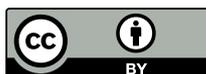

## 1 Introduction

The advent of the Internet of Things has brought us the *prosumer* [20, 34]. In software, a prosumer is a software component that both produces and consumes data, e.g., by producing and consuming distributed data streams. An example of a prosumer device is a modern smartwatch, which continuously consumes data (e.g., notifications, weather forecast) and is packed full of sensors that produce data (GPS, gyroscope, heart rate sensor, blood oxygen sensor, …).

Programmers face various demands when developing prosumer applications, which we characterise as *reactive*, *distributed*, and *open*. First, they are inherently *reactive* due to their highly interactive (event-driven) nature. Users expect them to update in real-time. Second, data may be produced and consumed by many *distributed* devices and sensors that are connected via a network (e.g., Bluetooth). Finally, we say that they are connected to *open* networks, meaning that devices and sensors continuously and spontaneously join and leave the application.

*Reactive programming* (RP) is an emerging approach to build reactive (prosumer) applications because it offers a solution to the problems of *callback hell* and *inversion of control* that arise when programming event-driven applications [2, 10]. Frameworks based on "reactive streams" such as ReactiveX [31] and Akka Streams [35] have been adopted into mainstream development, and functional RP is regularly proposed for building distributed systems [7, 24, 26, 27, 28, 33, 40].

In this paper we analyse the problems that occur when using an RP language or framework to build prosumer applications, i.e., distributed reactive applications for open networks. We identify two key areas where existing literature can cause issues. First, a strong *acquaintance discovery* mechanism is essential to discover prosumers as they join and leave the application. Such a suitable mechanism is rarely present in existing RP languages and frameworks. Second, as prosumer applications process events from *many* devices, an *acquaintance maintenance* mechanism in the underlying reactive program is essential. We found that existing RP languages and frameworks either exhibit a lot of accidental complexity, or are inefficient.

This paper is structured as follows. In Section 2 we analyse the problems that occur in state of the art RP languages and frameworks when implementing reactive prosumer applications. Due to their open nature, two different kinds of reactivity will be identified: *application-level reactivity* and *topology-level reactivity*. The running example throughout the rest of the paper is an app that counts bikes that roam around a city (Section 3). In Section 4 we introduce Stella, a reactive language whose *actors* and *reactors* form the basic building blocks of prosumer applications. In Section 5 we introduce *flocks*, a novel acquaintance discovery mechanism that supports the open nature of prosumer applications and that epitomises the aforementioned topology-level reactivity. In Section 6 we define a new distributed RP primitive called `deploy-*` that correctly and efficiently maintains the topology of a prosumer application based on sets of discovered prosumers. Finally, in Sections 7 and 8 we evaluate `deploy-*` qualitatively and via benchmarks, and in Section 9 we evaluate the distributed capabilities of flocks by running a bicycle simulation on a Raspberry Pi cluster.

## 2 State of the Art and Problem Statement

When using RP to implement prosumer applications (e.g., using ReactiveX [31], Akka Streams [35], REScala [37], ...), every prosumer in the network represents (part of) a reactive program. Prosumers interact with each other via *streams*, which are flows of events such as sensor measurements.

One of the main challenges we faced when implementing prosumer applications using RP is caused by the open network assumption. At every point in time, a prosumer application must know its set of *acquaintances* that can be reached over the network. These acquaintances vary throughout the lifetime of the application as prosumers join and become unreachable. Hence, a central aspect is *acquaintance management*: a mechanism to discover acquaintances, subscribe to their streams in order to react as they appear, and gracefully close the streams as they disappear.

In this section we explore acquaintance management for reactive programs, and we will show that existing RP languages and frameworks are either inefficient, or require a complex and error-prone mix of code when implementing prosumer applications.

### 2.1 Acquaintance Discovery: Extensional vs. Intensional

Since prosumers are not necessarily producers of services, we have replaced the traditional term "service discovery" by a more general term "acquaintance discovery" which comprises one prosumer discovering the existence of another prosumer. We classify acquaintance discovery mechanisms as either *extensional* or *intensional*. We use these terms in the traditional mathematical sense, namely that an *extensional definition* of a concept formulates its meaning by explicitly specifying every object that falls under the definition, and an *intensional definition* gives meaning to a concept by specifying the rules or conditions that objects have meet to be part of the definition. Applying these terms to acquaintance discovery:

**Extensional acquaintance discovery** means that the developer *explicitly* designates the acquaintances on the network. This is demonstrated by the following example in RxJS, a JavaScript implementation of ReactiveX [31, 36], which explicitly names the URL to a WebSocket echo server and binds the resulting stream to `echoStream`.

```
const echoStream = rxjs.webSocket.webSocket("wss://echo.websocket.org");
```

**Intensional acquaintance discovery** means that the developer writes a *prescription* of the acquaintances that can found on the network. An example can be found in the AmbientTalk/R reactive programming language [7], where the authors build a mobile peer-to-peer ticket trading application. The following snippet of their application discovers all tickets offered by ticket vendors on same (wireless) network. The expression `ambientBehavior:` creates a set of all objects on the network with the `TicketOfferT` type tag. Its result is a so-called *signal* (a reactive value) that automatically updates whenever the underlying collection updates.

```
def allNearbyOffers := ambientBehavior: TicketOfferT @All;
```

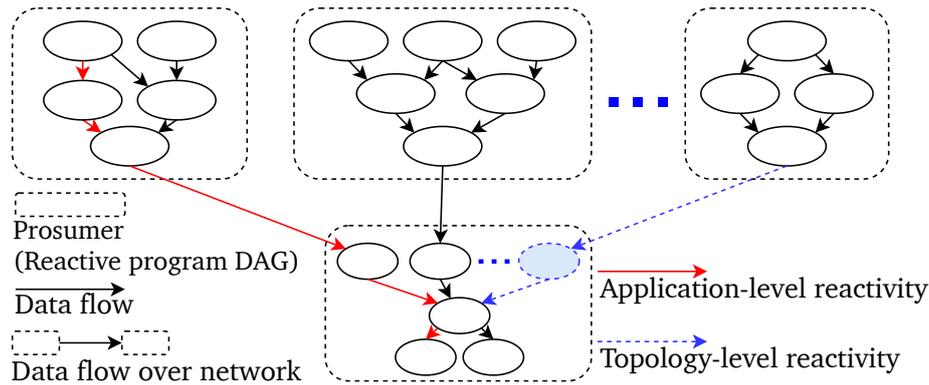

■ **Figure 1** Illustration of the 2 different levels of reactivity.

When dealing with *open* networks, an intensional discovery mechanism is essential: extensionally enumerating all possible acquaintances is by definition impossible. Moreover, because prosumers spontaneously come and go, the mechanism itself must also be reactive. However, existing RP languages and frameworks such as Flapjax [25], REScala [37], FS2 [14], Project Reactor [29], ReactiveX [31], and Streamz [41] support only extensional acquaintance discovery. The lack of built-in intensional acquaintance discovery mechanism leaves them in an unfavourable position. This is because each application needs to implement its own ad-hoc solution for intensional acquaintance discovery and to maintain its state as acquaintances appear and disappear, which leads to complex and error-prone code.

## 2.2 Acquaintance Maintenance in Reactive Programs

An overarching strategy is needed to correctly and efficiently *maintain* the state of the reactive program as acquaintances appear and disappear. For example, consider a reactive program that computes the average temperature of all thermometers that are connected to a network, and an acquaintance discovery mechanism that is capable of discovering them. Speaking in terms of streams, the `average` computation must appropriately react to streams appearing, disappearing, and updating with new values. In Figure 1 we illustrate these interactions between prosumers and their streams. Each of the rounded rectangles represents one prosumer, whose reactive program is represented by a Directed Acyclic Graph (DAG) that shows how data flows through the program [2]. At the top, Figure 1 depicts N (possibly different) producers of data, e.g., thermometers. At the bottom, a sole consumer reacts to prosumers and their data. We discriminate 2 *levels* of reactivity.

**Application-level reactivity:** Whenever an input node of a DAG changes (at the top of the DAG), the reactive language or framework automatically recomputes the dependent parts of the program that are affected by the change. We illustrate one such propagation path in red. We call this *application-level reactivity*, which equates to the "normal" propagation of values in existing work, e.g., temperature measurements that flow through the program.

■ **Listing 1** Topology-level reactivity in RxJS, calculating an average of all sensors.

```
1  const average$ = sensorDiscoveryService.sensors$.pipe(
2    zipWithIndex,
3    rxjs.flatMap(([sensor$, id]) => sensor$.pipe(
4      rxjs.map((tempr) => [id, tempr]),
5      rxjs.endWith([id, null]),
6      rxjs.catchError(_ => rxjs.of([id, null])))),
7    rxjs.scan((tracker, [id, tempr]) =>
8      tempr === null ? tracker.remove(id) : tracker.update(id, tempr),
9      new AverageTracker()),
10   rxjs.map(tracker => tracker.getAverage()));
```

**Topology-level reactivity:** A different kind of reactivity that we call *topology-level reactivity* occurs in consumers whose computations depend on an open number of producers, e.g., the average calculation. The topology of the DAG needs to be continuously reconfigured to accommodate the appearing or disappearing streams. This is illustrated in Figure 1 in blue, denoting the appearing and disappearing of a stream (and its dependencies) in the DAG of the consumer.

Application-level reactivity is well understood from languages and frameworks such as Fran [11], FrTime [9], REScala [37], ReactiveX [31] and Akka Streams [35]. In the remainder of this section we analyse how topology-level reactivity is handled by existing work on reactive programming. Here, there are two main approaches based on *discrete* and *continuous* representations of reactive values based on *event streams* and *behaviours* respectively [2]. Since they lead to a different programming style, we will discuss them separately in Sections 2.2.1 and 2.2.2.

### 2.2.1 Acquaintance Maintenance Using Event Streams

The mainstream approach to writing reactive programs is based on *discrete* reactive values, where the basic unit of change are events. Many RP languages and frameworks offer abstractions that represent event streams [9, 11, 25, 37], and frameworks based on *reactive streams* (i.e., event streams) are often used in mainstream software development [31][35].

The main ideas can be explained via an average computation that computes the average value of all thermometers connected to a network, which we implemented in RxJS [36], a state-of-the-art JavaScript streaming framework. Listing 1 implements this computation via an RxJS stream called average$ ($ is a naming convention for streams). To communicate the continuously appearing and disappearing sensors to the reactive program we used a stream called sensorDiscoveryService.sensors$ (not defined here) from which average$ is derived. This stream propagates a new sensor$ stream (containing temperature measurements) every time a sensor appears. In RxJS, the average$ stream is defined by "piping" the values from sensors$ through a sequence of RxJS stream operators. The operators in the example work as follows:

**Line 2, zipWithIndex:** Adds an identifier to each value of the sensors$ stream, yielding [sensor$, id] pairs where sensor$ is the aforementioned sensor's stream.

**Line 3, flatMap:** The lambda given as argument to flatMap is invoked for every [sensor$, id] pair and generates a new stream. flatMap remembers all streams it has generated and echoes their values on a first-in first-out basis. We use it to transform a stream of sensors to a stream of tuples [id, tempr] where id is the same as before, and tempr is its latest temperature measurement. Whenever the sensor$ stream is updated by said sensor, a new [id, tempr] pair is propagated by flatMap. Lines 5 and 6 handle the removing of a thermometer's stream (gracefully or via an error). They ensure that an [id, null] pair is propagated to "clean up" the thermometer downstream.

**Line 7, scan:** Most of the application logic is tackled by scan, that essentially implements a fold operation for streams where the accumulator is emitted for every input value. The accumulator is a purely functional AverageTracker (implemented elsewhere) that tracks the latest temperatures of each sensor. The subsequent map (line 10) extracts the current average.

The main problem with this approach is its accidental complexity, and we find that the code is also error-prone. In our experience this problem translates to other frameworks like Akka Streams as well, which operate in a very similar manner. Besides the essential complexity of the application logic (averaging temperatures), the operations needed to handle topology-level reactivity and to manage acquaintances (thermometers) are accidental complexity. In our experience finding the correct combination of operators is also difficult. The code snippet of Listing 1 was actually written by an anonymous reviewer of an earlier version of this paper. The original code snippet written by us contained a semantic bug, because we thought of the solution in a different way (more similar to the solution in Section 2.2.2). This code can still be found in Appendix A where we discuss the bug.

### 2.2.2 Acquaintance Maintenance Using Behaviours

Besides event streams, many RP languages and frameworks also feature continuous *behaviours*. Those that support both styles (e.g., Fran [11], FrTime [9], Flapjax [25] and REScala [37]) also include operations to convert one to the other, and either one can be used to implement the other [8, section 2.10]. Whereas event streams use operations such as map and filter, behaviours are programmed by lifting regular functions, leading to code that is typically more declarative and compact.

Listing 2 implements the same average calculation from the previous section in REScala [26, 37], a functional RP library in Scala. Line 1 declares a behaviour (called a signal in REScala) that holds values of type Set[Sensor]. Line 2 derives a new signal that computes their average value. The implementation is standard Scala: line 3 computes the sum total of all sensor values via a standard Scala foldLeft, and line 4 divides the total by the number of sensors. Note that each .value access automatically creates a dependency on a signal. In contrast to the RxJS code in Listing 1, the topology of the DAG that arises from this computation is automatically maintained by REScala.

Behaviour-based code is usually inefficient because of the propagation of (in this case) sets. Every time a change occurs at the topology-level *or* the application-level, a completely new set is automatically propagated through the reactive program, and all derived computations such as foldLeft are recomputed from scratch. Techniques

■ **Listing 2** Topology-level reactivity in REScala.

```
1  val sensors: Signal[Set[Sensor]] = … // implementation omitted for brevity
2  val average: Signal[Int] = Signal.dynamic {
3    val total = sensors.value.foldLeft(0) { (accum, sensor) => accum + sensor.measurement.value }
4    total / sensors.value.size }
```

such as *incremental data structures* [23] and *incremental behaviours* [32] have been proposed to speed up application-level reactivity. However, so far these techniques have not been integrated to speed up topology-level reactivity.

### 2.3 Summary and Problem Statement

To summarise the problems that we found in state of the art RP languages and frameworks for dealing with topology-level reactivity:

**Event streams**  result in code that reacts efficiently on both the application-level and the topology-level. However, event streams seem to engender larger and more complex code that is more difficult to write.

**Behaviours**  result in code that is more idiomatic and seems to be easier to write and understand [39], but is usually inefficient for both levels of reactivity.

In this paper we use a language-based approach to achieve code that is idiomatic (like behaviours) but also efficient. We present our solution using a programming language called Stella [43], which allows us to demonstrate and communicate, in our opinion, idiomatic code. Besides the code style, Stella implements a functional reactive programming model based on *deployments* of reactive code that will be an important part of the overall solution.

Intensional acquaintance discovery will be conceived via so-called *flocks*, a new abstraction to automatically discover prosumers such that *both* levels of reactivity can be correctly and efficiently implemented. A flock offers a stream to the reactive program that propagates a prosumer set, but instead of continuously propagating new sets, it will propagate *patches* that are akin to "join" and "leave" events. To correctly and efficiently maintain the state of the reactive program, we leverage these so-called *incremental* sets to implement a topology-reactive operator called deploy-*. The key feature of deploy-* is that it incrementally changes the topology of the DAG based on patches, and it meticulously patches its output collection whenever reactions occur on either the topology-level or the application-level.

## 3  Running Example: Whereabikes

Many cities around the world have infrastructure for shared bicycles or electric steps, which are increasingly network-enabled to track their location, battery level, distance travelled, etc [21]. The running example used in this paper is a reactive bike counter called "Whereabikes" that counts all bikes within a user-designated area. A screenshot

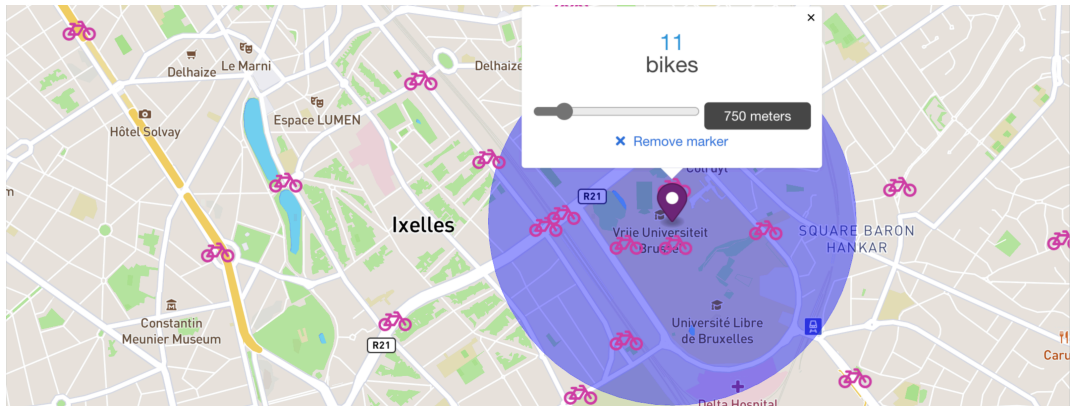

■ **Figure 2** Screenshot of the Whereabikes application (using mock bicycle data).

of Whereabikes is given in Figure 2, where a user placed a "counting marker" to count all bikes within a radius of 750 meters of our university campus. In the absence of real-world data, this implementation is completely user-controlled. I.e., a user mocks data via the GUI by manually adding, removing, and moving (via drag & drop) bikes and counting markers.

The example is conceptually simple, but challenging to implement. The programmer must ensure that bike counters correctly update in the following circumstances:

1. Bikes constantly move in and out of the designated area highlighted in blue.
2. Bikes spontaneously appear and disappear depending on network conditions, which is mocked by the user via GUI interactions.
3. A user may move the counting marker (drag & drop).
4. A user may change the radius wherein bikes are counted via a slider in the pop-up.

To give the reader a notion of scale, in 2019 *Villo!*, the public bike sharing program of Brussels, stationed 5000 bikes in the city dispersed over a total of 352 bike stations [30]. Consequently, the programmer must ensure that the computation is efficient.

## 4  A Brief Introduction to Stella

We use the reactive programming language called Stella [43] as a linguistic vehicle to express the core ideas of this paper[1] It is conceived as a continuation-passing style interpreter in TypeScript with trampolines and (green) threads [13, chapter 5]. Distributed peer-to-peer communication is supported via WebRTC [6].

Throughout this section we will explain code snippets that implement parts of the running example, and we detail the new concept of *reactor deployments* which we consider to be an important part of our solution. Throughout all Stella code snippets

---

[1] Some keywords were renamed since the version of Stella presented in [43], but the language semantics remain the same. E.g., actor was renamed to `def-actor`, reactor to `def-reactor`, and expressions with side-effects end with a "!".

■ **Listing 3**  Examples of basic expressions in Stella

```
1  (def a 1)                                        // (def <identifier> <expression>)
2  (set! a 2)                                       // (set! <identifier> <expression>)
3  (if (equal? a 1) (println! "y") (println! "n"))  // (if <condition> <consequent> <alternative>)
4  (def p (new LngLat 'at 4.39513130 50.8220238))   // (new <class> <constructor name> <...args>)
```

■ **Listing 4**  A "Hello World!" program in Stella.

```
1  (def-actor Main
2    (def-constructor (start env)
3      (println! "Hello World!")))
```

we highlight keywords in blue, strings are green and are surrounded by double quotes, and symbols start with a single quote and are highlighted in red.

### 4.1 Base Language and "Hello World!"

Stella is a dynamically typed language where all run-time values are objects. The language can be thought of as having two layers: a sequential object-oriented base language, and the concurrent level of *actors* and *reactors*. The sequential base language contains objects such as numbers, strings (e.g., "hello") and symbols (e.g., 'hello), as well as method invocations on objects and a number of special forms (e.g., to spawn actors and reactors).

Stella uses S-expression syntax, with operators in prefix notation. Dynamic method lookup in Stella's object-oriented model is single dispatch. The expression (println! "Hello World!") looks up the println! method on the class String of the receiver object "Hello World!". Similarly, the expression (+ 1 2) invokes + from the Number class on the receiver object, number 1, passing the number 2 as the argument. Some other examples from the base language are shown in Listing 3. Local variables are introduced via def, assignments use set!, conditionals use if, and classes are instantiated to objects via new (e.g., LngLat represents a GPS coordinate and has a constructor called at).

A program contains 4 types of top-level definitions: classes, *actor behaviours* ("the class of an actor"), *reactor behaviours* ("the class of a reactor", i.e., a reactive program), and *flocks*. The interaction between those last three concepts form the core of this paper. Starting a program is exemplified by the "*Hello World!*" program in Listing 4. A developer must define a Main actor behaviour that defines a constructor named start with 1 formal parameter (that contains environment variables). The body of its start constructor contains base language expressions, in this case a single println!.

### 4.2 Actors and Streams

An actor is a process that has a single *behaviour* and a mailbox [22]. The actor behaviour describes the internal state and interface of an actor. There is no shared state between actors (and *reactors*), and all messages between them are passed via deep copy. An

■ **Listing 5**  A simple "digital twin" for a bicycle in Whereabikes

```
1  (def-actor Bike
2    (def-stream location)
3    (def-constructor (init initial-location)          (emit! location initial-location))
4    (def-method (update-location! new-location)  (emit! location new-location)))
```

atypical feature of actors in Stella is that they can export streams to which they emit values. These serve as the basis for flocks in Section 5.

Consider the code in Listing 5, which implements a part of Whereabikes. We use a Bike actor behaviour to create a "digital twin" for every bicycle. While these actors are conceptually running on the physical bikes that move around a city, in our implementation they are spawned locally in response to GUI events. There are 3 declarations in its body. Line 2 declares a stream called location, which means that every actor with the Bike behaviour exports a stream called "location". Line 3 declares a constructor called init (there can be multiple), and line 4 declares a method called update-location!. Both have 1 formal parameter and an emit! statement in their body that emits a location (a GPS coordinate object) to the location stream of the current actor. Emitting to streams of other actors (and reactors) is not possible.

A stream can be seen as an event source, and an emission is an event publication. Emitting a value sends an asynchronous message to all actors (and reactors) that are subscribed to the stream. A reference to a specific actor's stream is obtained via a *qualification* expression using dot-notation. For instance, given an actor bike, the expression bike.location evaluates to an object of type Stream, namely the location stream of bike. Note that dot-notation is exclusively used to address streams, and not fields, methods, etc.

Actors can be spawned via spawn-actor! in the base language (e.g., in the Main actor). As shown below, it accepts at least 2 arguments: the name of an actor behaviour (e.g., Bike) and a symbol denoting the name of its constructor (e.g., init). Any other arguments are passed to the constructor, in this case a LngLat object. An ActorReference object is returned that is used to send asynchronous messages to the actor.

```
(def bike (spawn-actor! Bike 'init (new LngLat 'at 4.39513130 50.8220238)))
```

### 4.3  Monitoring Streams

Actors can monitor streams for emissions using monitor!, as demonstrated in Listing 6, where the Main actor spawns a new Bike and monitors its location stream (line 4). Every time the location stream emits a new value, the Main actor will receive a log! message in its mailbox. When this message is eventually dequeued and processed, the Main actor invokes the corresponding log! method with the emitted value as argument.

■ **Listing 6**   Monitoring a bike's location with an actor.

```
1  (def-actor Main
2    (def-constructor (start env)
3      (def bike (spawn-actor! Bike 'init (new LngLat 'at 4.39513130 50.8220238)))
4      (monitor! bike.location 'log!))
5    (def-method (log! lnglat) (println! "bike pos: " lnglat)))
```

### 4.4  Functional Reactive Programming With Reactors

Besides actors, Stella features *reactors*. A reactor is a process that encapsulates a functional reactive program [2]. In general, the fundamental mechanism of functional reactive programming can be easily explained using the programming model of spreadsheets. When a cell C1 contains the expression A1 + B1, then the value of C1 is automatically recomputed every time the contents of cell A1 or B1 changes. Behaviour-based RP languages (see Section 2.2.2) operate on the same principle, but elevated to the level of programming languages. By encapsulating imperative code in actors and reactive code in reactors, Stella avoids the issues that arise when combining imperative and reactive programming. Stella's object model ensures that reactors are purely functional, and that their computations always terminate [43].

A reactor contains one or more *reactor deployments*, which are "instances" of a *reactor behaviour*, each with their own run-time state. Reactor behaviours, deployments, and reactors themselves are explained in the remainder of this section.

#### 4.4.1  Reactor Behaviour

Just like an actor is spawned from an actor behaviour, every reactor is spawned from a *reactor behaviour*. A reactor behaviour has a name, at least one *source* (its "input"), at least one *sink* (its "output"), and any number of expressions that describe the computations between the sources and sinks. For example, the reactor behaviour called DegreesToRadians in Figure 3a converts a mathematical angle in degrees to radians. Its single source is called degrees, and its single sink is denoted by out (the result of a multiplication). When DegreesToRadians is spawned (creating a reactor), the value of its sink will be automatically recomputed (just like a spreadsheet) whenever the value of degrees changes via incoming messages.

It is common to draw a reactive program as a Directed Acyclic Graph (DAG) which may aid the understanding of how data flows through the program [3, 38]. The corresponding DAG representation of DegreesToRadians is given in Figure 3b. Constant expressions such as the division are computed at compile-time and are collapsed into one DAG node. We often think about reactive programs as the propagation of values through the DAG from top to bottom, recomputing the expressions in the nodes along the way.

#### 4.4.2  Reactor Deployments

A *reactor deployment* is an "instance" of a reactor behaviour, of which there can be many within the same reactor. Each deployment can have different input values,

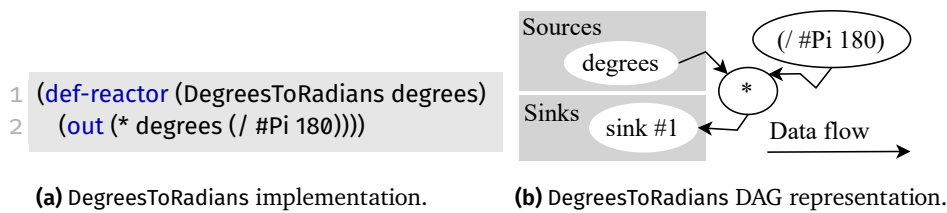

```
1  (def-reactor (DegreesToRadians degrees)
2    (out (* degrees (/ #Pi 180))))
```

**(a)** `DegreesToRadians` implementation.

**(b)** `DegreesToRadians` DAG representation.

■ **Figure 3**  Implementation and DAG of the `DegreesToRadians` reactor behaviour.

■ **Listing 7**  Deploying additional reactor behaviours.

```
1  (def-reactor (DistanceBetween point1 point2)
2    (def lat1 (get-lat point1))
3    (def lon1 (get-lng point1))
4    (def lat2 (get-lat point2))
5    (def lon2 (get-lng point2))
6
7    (def dlat (deploy DegreesToRadians (- lat2 lat1)))
8    (def dlon (deploy DegreesToRadians (- lon2 lon1)))
9    (def a (+ (expt (sin (/ dlat 2)) 2) (* (cos (deploy DegreesToRadians lat1)) (cos (deploy
         ↪ DegreesToRadians lat2)) (expt (sin (/ dlon 2)) 2))))
10   (def c (* 2 (atan2 (sqrt a) (sqrt (- 1 a)))))
11   (out (* 6367 c)))
```

output values, and run-time state. Deployments will serve as the basis for efficient application-level and topology-level reactivity.

Consider the `DistanceBetween` reactor behaviour in Listing 7 that calculates the great-circle distance between two points using the Haversine Formula [12]. Most expressions in its body are variable definitions and method invocations on objects. Since the formula requires some angles to be specified in radians, `DistanceBetween` makes use of the aforementioned `DegreesToRadians` behaviour by deploying instances of it. Responsible for this is the deploy expression, for example on lines 7 and 8.

The deploy operator can be seen as a function application for reactor behaviours, but instead of applying a function once, it deploys a reactor behaviour that stays active (continuously "waiting" for new source values to react to). Whenever the value of one of the input arguments changes, then this new value will also propagate through the corresponding deployment of `DegreesToRadians`. Similarly, the value of `DegreesToRadians`'s sink is used by the deployment of `DistanceBetween`, and will cause any dependent expressions to change as well.

**Qualification**  Continuing the analogy that a deployment is like a function application (but for reactor behaviours), a crucial difference is that value of the sinks of a deployment can change even when its given input arguments did *not* change. This is because deployments may create additional dependencies on streams exported by (re)actors that are propagated through the deployment. Consider `IsBikeWithinRadius` reactor behaviour in Listing 8 that checks whether one bicycle is within a certain distance of a given point, i.e., within the blue circle of Figure 2. Its inputs are a point

■ **Listing 8** Creating dependencies to streams within reactors.

```
1  (def-reactor (IsBikeWithinRadius point radius-meters bike)
2    (def distance-km (deploy DistanceBetween point bike.location))
3    (out (if (< (* distance-km 1000) radius-meters)
4             bike)))
```

■ **Listing 9** Supplying input values to a reactor.

```
1  (def-actor Main
2    (def-constructor (start env)
3      (def deg2rads (spawn-reactor! DegreesToRadians))
4      (react-to! deg2rads 4.39513130)
5      (monitor! deg2rads.out 'log!))
6    (def-method (log! rad) (println! "radians: " rad)))
```

(GPS coordinate), a radius in meters, and one bicycle (a reference to a Bike actor). The body calculates the distance between the point and the current location of the bike, and outputs the bike if the distance is smaller than the given radius.

Recall from Section 4.2 that the location of a bike was conceived as a location stream exported by a Bike actor. To calculate the distance to a bike, IsBikeWithinRadius creates a dependency on this stream via the bike.location qualification expression on line 2. Whenever such a qualification is used in a reactor, a subscription to the stream is established automatically, and the reactive program automatically updates whenever the bike's location changes.

### 4.4.3 Reactors

A *reactor* is a process with a mailbox and a set of reactor deployments. The first deployment, which we call the *root* deployment, is created when a reactor is spawned via a spawn-reactor! statement in the base language. A reactor continuously dequeues messages from its mailbox and propagates them through the corresponding deployment. After a full propagation turn, the values of the root deployment's sinks are emitted to the out stream exported by the reactor.

The philosophy of reactors is different from actors, because reactors themselves do not independently perform *actions*, and instead they can only *react* to input values. Whenever a reactor is spawned, it indefinitely waits for new values on its sources. For example, the Main actor in Listing 9 spawns a DegreesToRadians reactor from Figure 3a. Then, the actor usually "kick-starts" the reactor by changing the value of its sources using react-to! (communicated via an asynchronous message). For example, since DegreesToRadians has 1 source, line 4 instructs this source to change to the number 4.39513130. The actor may subsequently monitor the reactor's out stream to receive the corresponding value in radians.

### 4.5 Technicalities

When implementing reactors using a conventional DAG propagation algorithm using a priority queue, there can be *glitches* between deployments, which are temporary inconsistencies in the computations of a DAG. In Appendix B we explain how to prevent such glitches.

## 5  Intensional Acquaintance Discovery With Flocks

In this section we propose the key idea developed in the paper, namely an acquaintance discovery abstraction called a *flock*. A flock is a special type of actor that facilitates acquaintance discovery. In essence, a flock describes a collection of (re)actor references that are automatically shared over the network with other flocks that have the same name. Our definition of flocks was inspired by *volatile sets* [16] and flocks for ambient-oriented programming [5]. Both offer intensional acquaintance discovery in the same spirit as our flocks, but they have not been adapted to reactive programming.

Flocks are defined in top-level scope with a unique name. For example, the following code snippet defines a flock called Bikes.

```
(def-flock Bikes)
```

### 5.1 Publishing and Unpublishing Actors and Reactors

Local (re)actor references can be published to the network by adding them to a flock via publish!. For example, the devices that are running Bike actors from Listing 5 (Section 4.2) may (un)publish these actors via the following expressions:

```
1 (def bike (spawn-actor! Bike 'init (new LngLat 'at 4.39513130 50.8220238)))
2 (publish! Bikes bike)      // publish bike actor to the Bikes flock
3 (unpublish! Bikes bike)  // remove bike actor from the Bikes flock
```

### 5.2 Reading the Contents of a Flock

Every flock has a stream called contents, which is the primary means to access its continuously evolving acquaintances. Thus, the Stella expression Bikes.contents denotes a reference to this stream (see Section 4.2). The first value emitted by the flock to its contents stream is always a *snapshot*, and all subsequent values are a *patch*. We use these terms to more easily describe the use of a special IncrementalBag in combination with streams (inspired by [23, 32]).

**Snapshot:** The snapshot of a flock is an immutable IncrementalBag data structure (a set that may contain duplicate values) that records the contents of a flock at a specific moment in time.

**Patch:** A patch is emitted whenever the contents of a flock changes. A patch is an insert, update or remove object that can be applied to a snapshot.

Snapshots and patches are used to obtain efficient application-level and topology-level reactivity. The semantics of snapshots and patches align with regular Stella semantics for streams, where any (re)actor that subscribes to a stream immediately receives the most recently published value. For flocks this is always an up-to-date snapshot, rather than the latest patch.

### 5.3 Distributed Flocks

A defining property of flocks is that their contents are automatically shared over the network with all other flocks of the same type. Any additions via publish! or removals via unpublish! are automatically propagated to the other flock actors on the network, who incorporate those changes into their own contents. References to local (re)actors are automatically transformed to remote references when they are passed over the network. In the event of a peer or network failure, each peer individually determines the (re)actors that can no longer be reached, which are then automatically removed from that peer's flock. Whenever a disconnected peer rejoins the network, the removed (re)actors are re-added to the flock. This process of synchronising state and monitoring the network is always running in the background for every flock, hence why they are explicitly conceived as an actor.

Our current implementation of Stella makes use of a discovery server to facilitate the discovery of the Stella interpreters running on different peers in the network. A signalling (discovery) server is required by WebRTC to exchange initial peer information without user interaction. Afterwards all communication is directly peer-to-peer [1]. The discovery server is also responsible for keeping a total overview of the complete contents of a flock, i.e., all (re)actors that a peer *should* be able to reach via the network. Every peer still independently determines the (re)actors that it can *actually* reach. We implemented this via heartbeats between all peers that share a flock.

Currently, any other Stella program can publish (re)actors to a flock. From a security point of view, we consider an authentication mechanism (e.g., to discover only authorised (re)actors) to be an orthogonal concern. Such mechanisms can be implemented on the meta-level (e.g., on the level of socket connections), via a type system, etc. At the time a (re)actor is included in a flock, we already assume that the other parties are trusted.

## 6  Efficient Application-Level and Topology-Level Reactivity

We show how reactors are used in combination with reactor deployments (cf. Section 4.4.2) and flocks (cf. Section 5) to support efficient application-level and topology-level reactivity. We introduce an efficient topology-reactive operator for reactors called deploy-*, and we demonstrate its semantics by implementing the bike counters from the running example.

■ **Listing 10**  Counting all bikes within a radius.

```
1  (def-reactor (CountingMarker id location radius)
2    (def all-bikes Bikes.contents)
3    (def bikes-nearby (deploy-* (bind IsBikeWithinRadius location radius) all-bikes))
4    (out id (size bikes-nearby)))
```

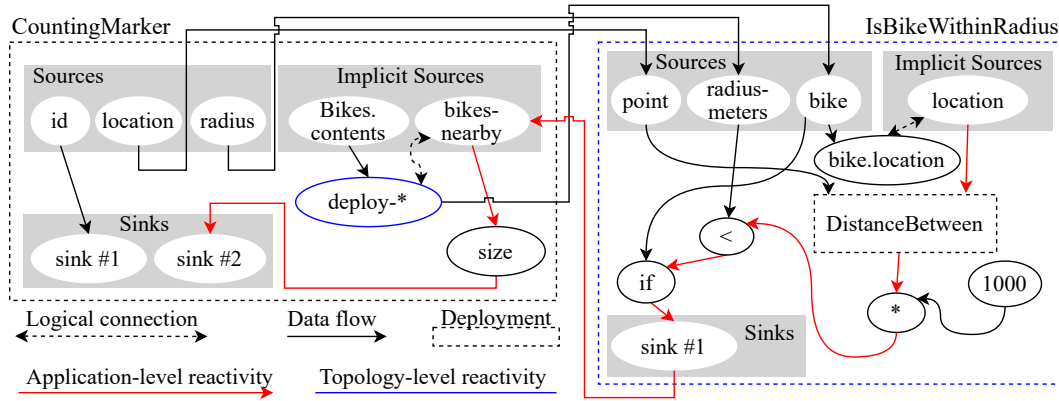

■ **Figure 4**  CountingMarker DAG representation (Listing 10).

## 6.1 Topology-Level Reactivity: "deploy-*"

Listing 10 implements a reactor behaviour called CountingMarker that counts all bikes within an area. It has 3 sources: id is a unique identifier for the marker, location is the centre of the circular area wherein bikes are counted, and radius determines the radius (in meters) of the area. Its body is structured as follows. Line 2 defines a variable all-bikes that refers to the Bikes.contents stream, which emits all bikes that can be discovered on the network (a snapshot followed by patches). Line 3 defines bikes-nearby as the result of a deploy-* operation, which conceptually transforms a snapshot of all bikes to a snapshot that contains only the bikes that fall within the given radius. Line 4 declares 2 sinks, namely the id of the marker and the size of the bikes-nearby collection. These 2 values are published together on the out stream of the reactor whenever either of them changes.

The deploy-* operator can be thought of as a regular "map" operation for lists, but instead of mapping a function over a list, deploy-* deploys ("instantiates") a new reactor behaviour for every element in the given collection. Its first argument is the reactor behaviour to "map" over the collection, in this case given by a bind expression. The bind operator implements partial application for reactor behaviours (not to be confused with monadic bind). The first two sources of IsBikeWithinRadius (cf. Listing 8) will depend on location and radius, and its third source remains "unconnected". The second argument of deploy-* is a collection given by all-bikes. At run-time, a new deployment of IsBikeWithinRadius is created for each bike in the collection (in no particular order), and the corresponding bike is propagated via the sole unconnected source of the deployment. The result of deploy-* is a collection that contains the result of each created deployment, i.e., the value of their sinks (of which there must be exactly 1 for each deployment).

The creation of deployments is visualised in Figure 4, which depicts the DAG representation of `CountingMarker`. Here, the blue `deploy-*` node has created one deployment of `IsBikeWithinRadius` whose sources are connected to the corresponding nodes in the `CountingMarker` deployment. The sink node of `IsBikeWithinRadius` is connected to the `bikes-nearby` implicit source. It is called "implicit" because it is generated by the DAG compiler to receive values from outside the current deployment, and in the case of `deploy-*`, it also constructs the correct output.

Technically, `deploy-*` offers "filter-map" semantics where values that signify "no value" (e.g., `#undefined`) are excluded from the result. This is the case in our example, since the implementation of `IsBikeWithinRadius` (Listing 8) did not provide an alternative branch for the if-test. When this is not desirable, `deploy-*` accepts an optional third argument (not used here) to replace those empty values with a default.

## 6.2 Incremental Updates

Efficient application-level and topology-level reactivity is achieved by meticulously propagating snapshots and patches whenever a change occurs. For example:

**Topology-level patching:** Whenever the `Bikes` flock emits an `insert` patch that adds a bike, then `deploy-*` also propagates an `insert` patch with the output of the newly created `IsBikeWithinRadius` deployment.

**Application-level patching:** Whenever a bike is within the counting radius and suddenly its location updates to being outside the counting radius, then `deploy-*` propagates a `remove` patch that removes the bike from the output.

We have devised a set of rules that show how `deploy-*` transforms its input to output, i.e., which snapshots or patches are propagated as output in response to a particular change in the input. As reactive programs are conventionally drawn as a DAG, we visualise these rules as a diagram that corresponds to a part of the DAG. Their structure can be explained using Figure 5a:

- Arrows denote the direction in which data flows, i.e., from left to right.
- Ellipses correspond to nodes in the DAG. The left ellipse corresponds to the main `deploy-*` node (the blue node in Figure 4), and the right ellipse corresponds to its implicit source node (called `bikes-nearby` in Figure 4).
- The input value of `deploy-*` is shown in the top left, and its output in the top right. All values $v$ within snapshots and patches are associated with a *deployment key* $d$, denoted as $d \rightarrow v$. A deployment key is a unique identifier generated by the implementation of a snapshot (an `IncrementalBag`) when a value is inserted.
- Reactor deployments are drawn as rectangles and labelled with a deployment key that identifies the deployment.
- The run-time state of `deploy-*` is a snapshot. In the bottom left we show the pre-snapshot, i.e., the snapshot before the new input value has propagated through the DAG. The post-snapshot is given in the bottom right, i.e., `deploy-*`'s new state after propagating the input value.

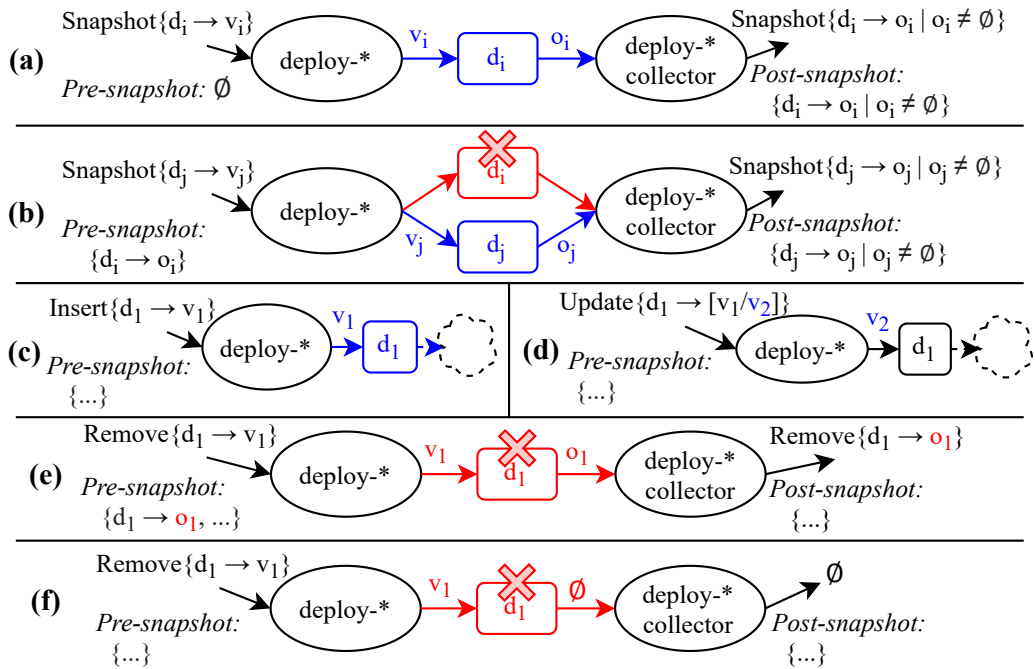

■ **Figure 5** Topology-level reactivity rules.

### 6.2.1 Efficient Topology-Level Reactivity

**Snapshot (Figure 5a):** The first value propagated by flocks is always a snapshot. This propagation is depicted in Figure 5a where the pre-snapshot is empty. A new deployment $d_i$ is created for each deployment key $d_1 \ldots d_i$ in the input snapshot, and the corresponding value $v_1 \ldots v_i$ is propagated through the corresponding deployment. The outputs $o_i$ are collected into a new snapshot with the same deployment key. Since deployments can produce "no value" (e.g., `#undefined`), those values $\emptyset$ are omitted from the output and post-snapshot.

**Snapshot replace (Figure 5b):** While the situation does not arise when using flocks, in general the input of `deploy-*` can be replaced with an entirely new snapshot. In this case all existing deployments $d_i$ from the pre-snapshot are replaced with new deployments $d_j$. The output is a new snapshot containing the outputs of the new deployments $o_j$, and the post-snapshot is adjusted accordingly.

**Patch insert (Figure 5c):** An insert patch adds a new deployment key $d_1$ with value $v_1$. A new deployment $d_1$ is created (highlighted in blue) with the new value $v_1$ as input. We omit the generated output and post-snapshot, denoted by the dashed cloud, as these follow the same rules as application-level reactivity (Section 6.2.2).

**Patch update (Figure 5d):** An `update` patch for a deployment key $d_1$ replaces its old value $v_1$ with a new value $v_2$. The new value $v_2$ is propagated through the existing deployment $d_1$. The generated output and post-snapshot follow the same rules as application-level reactivity (Section 6.2.2).

**Patch remove (Figure 5e):** To remove a value $v_1$, the corresponding deployment $d_1$ is removed. Since the pre-snapshot denotes that deployment $d_1$ previously contributed a value $o_1$ to the output, the new output is a remove patch with deployment key $d_1$ and value $o_1$.

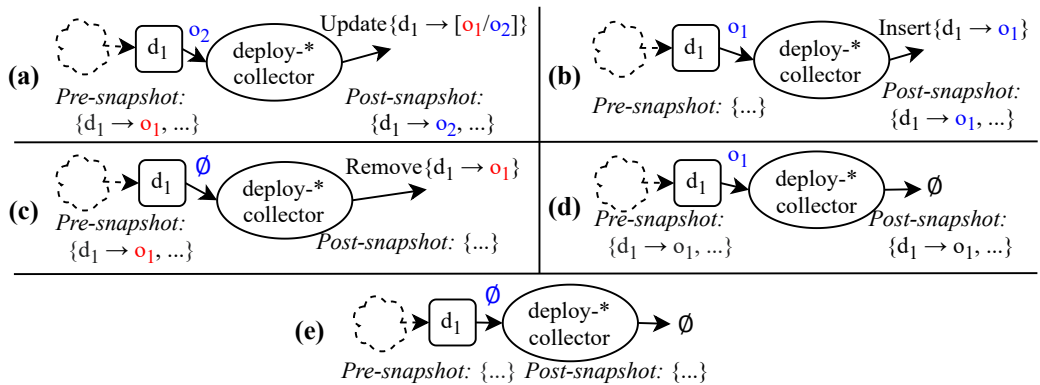

**Figure 6** Application-level reactivity patch rules.

**Patch remove* (Figure 5f):** A corner case of a remove patch is when the deployment $d_1$ did not contribute a value to the pre-snapshot. In this case no new remove patch is propagated as output (i.e., the old result remains valid).

### 6.2.2 Efficient Application-Level Reactivity

The set of rules in Figure 6 show the output that is generated after a reactor deployment produces a new value. Each diagram omits the left part (denoted by a dashed cloud).

**Production (Figure 6a):** A deployment $d_1$ propagates a new value $o_2$, and $d_1$ is already associated with the output $o_1$ in the pre-snapshot. In this case, propagate an update patch that replaces value $o_1$ with $o_2$ and update the post-snapshot.

**Sudden production (Figure 6b):** After a period of producing no value $\emptyset$, a deployment $d_1$ suddenly propagates a new value $o_1$. In this case, propagate an insert patch for $o_1$, and add this value to the post-snapshot. This situation arises, for example, when $o_1$ is the first value propagated by the deployment $d_1$ after its creation.

**Remove production (Figure 6c):** After a period of producing values, a deployment $d_1$ suddenly produces "no value" (e.g. `#undefined`). In this case, propagate a remove patch that contains $o_1$ and remove $o_1$ from the post-snapshot.

**Equal production (Figure 6d):** Since reactive programs are purely functional, typically only distinct values are propagated. Thus, if a deployment $d_1$ produces the same value $o_1$ included in the pre-snapshot, then no patch is propagated.

**No production (Figure 6e):** When a deployment $d_1$ produces "no value", and $d_1$ is not included in the pre-snapshot, then propagate nothing.

### 6.3 Discussion

There are a number of concerns and drawbacks related to our approach that should be taken into account.

**Method calls:** When propagating patches, all operations that are executed on snapshots are automatically "patch-aware". For example, the invocation of size on bikes-nearby

■ **Listing 11** The AddAll behaviour accumulates all values in an IncrementalBag.

```
1  (def-reactor (AddAll snapshot-of-numbers)
2    (out (fold snapshot-of-numbers 0 '+ '-)))
```

(Listing 10 line 4) automatically takes into account patches, such that previously computed results can be incrementally adapted. Some methods, such as a fold, require information about the values that were removed and replaced, hence why these values are included in update and remove patches.

**Supported data structures:** In principle any iterable collection can be provided as input to deploy-*, but not all types are equally suitable. Stella currently supports a Bag and IncrementalBag. Particularly an IncrementalBag lends itself to an $O(1)$ implementation of deploy-* for every insert, update, or remove patch. Other data structures such as lists and vectors are likely to introduce extra algorithmic complexity and computational overhead to the implementation of deploy-*, specifically because extra work is required to create and maintain order in the output collection [23], e.g., making sure indices in a vector are sequential and without gaps.

**Developer convenience:** Unfortunately snapshots and patches are not entirely transparent, and programmers have to be aware of them. For example, consider the AddAll reactor behaviour in Listing 11 that sums all numbers in a snapshot. The fold method in the interface of IncrementalBag efficiently aggregates its values. However, its expected arguments differ from a conventional fold: besides an initial accumulator and the + method name to add values to the accumulator, developers must also provide an operation to "reverse" the accumulator, in this case - (minus) for numbers. Otherwise it is not possible to efficiently update the accumulator after an update or remove patch.

## 7  Qualitative Evaluation: Comparison to State of the Art

In Section 2.2 we implemented an average computation that calculates the average temperature of a set of thermometers in two styles using reactive streams (in RxJS) and functional RP (in REScala). We argued the RxJS code exhibits accidental complexity, whereas the REScala code is more idiomatic, but inefficient. Our goal with Stella was to implement the example using idiomatic code like functional RP (Section 2.2.2), but also efficient. In this section we show an implementation of the same example in Stella. In Section 8 we will evaluate deploy-*'s performance.

Listing 12 implements the aforementioned average computation. It consists of a Thermometers flock (line 1), a SensorValue reactor behaviour (line 2), and the main logic which is implemented by the Average reactor behaviour (line 3). In a nutshell, deploy-* is used to transform a snapshot of thermometers to a snapshot of their latest measurements, which are subsequently averaged using the fold operation discussed in Section 6.3. Stella's programming style is more similar to the behaviour-based REScala code.

■ **Listing 12**  Averaging the temperature of a set of thermometers in Stella.

```
1 (def-flock Thermometers)
2 (def-reactor (SensorValue sensor) (out sensor.value))
3 (def-reactor (Average)
4   (def measurements (deploy-* SensorValue Thermometers.contents))
5   (def sum (fold measurements 0 '+ '-))
6   (out (/ sum (size measurements))))
```

Compared to RxJS, deploy-* seems to be more restricted in use than RxJS' flatMap (used in Listing 1) and its siblings (e.g., switchMap). But these restrictions are no accident: the features and restrictions of deploy-* are due to it being tailored specifically for handling topology-level reactivity, whereas flatMap and its siblings operate at a broader level of abstraction. Hence, it is more difficult to use them for topology-level reactivity, e.g., when dealing with disappearing information (e.g., disconnecting thermometers).

The main difference between Stella's code and the REScala code from Listing 2 is that topology-level reactivity is *explicitly* managed via deploy-*, whereas in REScala it is implicitly managed as a consequence of using REScala in combination with Scala's fold. Using Stella's approach, efficient topology-level reactivity is achieved at the cost of one single line of code.

## 8 Evaluation: Algorithmic Complexity

We provide experimental evidence that our approach leads to efficient application-level *and* topology-level reactivity. Since raw (real-time) performance is not a goal of Stella, it is difficult to compare raw performance to existing systems (e.g., REScala). Therefore the benchmarks compare the algorithmic complexity of our approach and of the typical approach taken by state of the art RP languages such as REScala, but implemented using Stella.

### 8.1 System and Benchmarking Specifications

Experiments were run on Ubuntu 20.04.2 LTS and Node.js v14.16.0, with the command-line options –trace-gc –max-old-space-size=65536. While Stella is single-threaded, experiments were run on an AMD Ryzen™ Threadripper™ 3990X with 128GB of DDR4-3200 RAM, of which 64GB were available to Node.js (abundant for our application). We used Node.js' "performance measurement API", which implements the W3C recommendation for high resolution time with sub-millisecond precision [15].

### 8.2 Benchmarking Application-Level Reactivity

We measure the time it takes (on average) for a value to propagate through a CountingMarker reactor from Listing 10. To this end, we ran two sets of experiments to

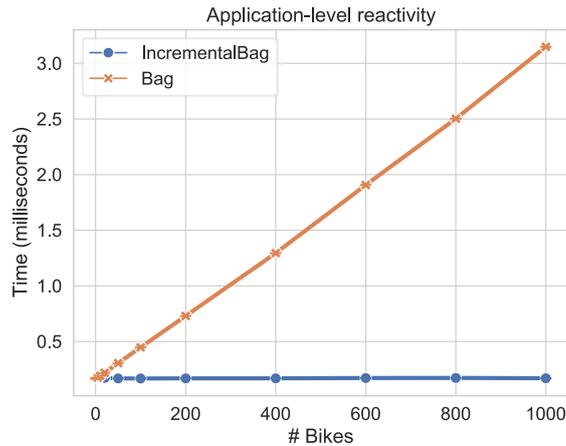

■ **Figure 7** Application-level reactivity (propagating one location update). Error bars = 99% confidence interval (drawn, but barely visible due to their small size).

propagate a varying number of bikes (from 1 to 1000) using an `IncrementalBag` and a (non-incremental) `Bag`. The `Bag` experiments replicate the algorithmic complexity of the same program implemented in an existing behaviour-based RP language such as REScala. Then, we measure the time it takes for the `CountingMarker` reactor to process 1 location update from a bike. To obtain statistically reliable results and to exceed any internal VM thresholds for optimisation [4], we measured 100,000 updates evenly spread over all bikes in the experiment. Total benchmark run-time ranged between 22 seconds (1 bike) and 6 minutes (1000 bikes).

If our performance claims hold and if our proposed mechanisms are correct, then the measured execution time will grow with the number of bikes when using a `Bag`, but will remain constant when using an `IncrementalBag`.

**Results**    Based on a manual inspection of the data (via a scatterplot), between 50-500 updates are required by the JavaScript VM (V8) to warm up, so we removed the first 500 measurements for each experiment from the compiled results shown in Figure 7. The graph shows that the run-time using a `Bag` grows linearly with the number of bikes. This is expected, since every time the location of one bike is updated, `deploy-*` constructs and propagates a new `Bag` (which is the case in existing behaviour-based RP literature such as the fold in Listing 2). In contrast, when using an `IncrementalBag`, run-times remain constant as the number of bikes increases because only a patch is propagated. Note that we draw error bars denoting a 99% confidence interval, but they are barely visible due to their small size.

## 8.3   Benchmarking Topology-Level Reactivity

We measure the time it takes (on average) to modify the topology of the DAG of the `CountingMarker` reactor (Listing 10) when the number of bikes increases or decreases. Similar to the application-level reactivity experiments, we ran two sets of experiments

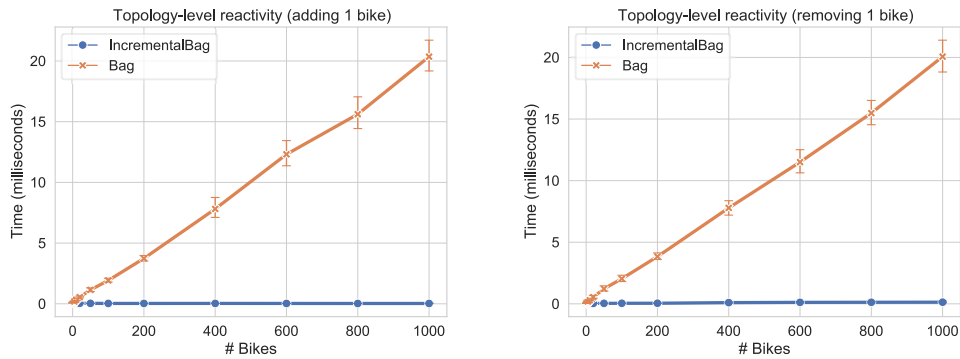

**(a)** Run-time cost of adding 1 bike.   **(b)** Run-time cost of removing 1 bike.

■ **Figure 8**  Topology-level reactivity (adding or removing 1 bike): comparison of the `Bag` and `IncrementalBag` experiments. Error bars = 99% confidence interval.

using an `IncrementalBag` and `Bag` with a varying number of bikes (from 1 to 1000). The `Bag` experiments replicate existing behaviour-based RP implementations.

Each experiment starts from a (fixed) number of `N` bikes (between 1 and 1000). First we propagate `N-1` bikes through the `CountingMarker` reactor (not measured). Then, we measure the time it takes for `deploy-*` to change the topology of the DAG when adding or removing the $N^{th}$ bike.

If our performance claims hold, then the experiments using a `Bag` exhibit longer execution times, because modifying the topology of the DAG involves more work. At the very least this requires a complete traversal over the `Bag` (i.e., $O(n)$).

To obtain statistically reliable results, we repeated each addition and removal of the $N^{th}$ bike 500 times, except for the `Bag` experiments where `N = {400,600,800,1000}`, which were repeated 100 times to reduce total run-time. Total run-time ranged between 2 seconds (1 bike) and 4 hours (1000 bikes). Note that the run-time is high, because there can be a lot of application-level reactivity work between consecutive measurements. This is already an indication that inefficient topology-level reactivity can have compounding consequences for total application run-time.

**Results**   Based on a manual inspection of the data (via a scatterplot) we determined that, depending on the experiment (`Bag` or `IncrementalBag`) and number of bikes, between 25-100 updates are required by the JavaScript VM (V8) to warm up. We removed the first 100 repetitions for all experiments with 500 runs, and the first 25 for the `Bag` experiments with 100 runs (with `N = {400,600,800,1000}`).

We compiled the measurements of adding 1 bike and removing 1 bike in separate graphs, shown in Figures 8a and 8b respectively. Both show that the run-time of `deploy-*` in the `Bag` experiments grows linearly with the number of bikes, whereas it remains constant for `IncrementalBag`.

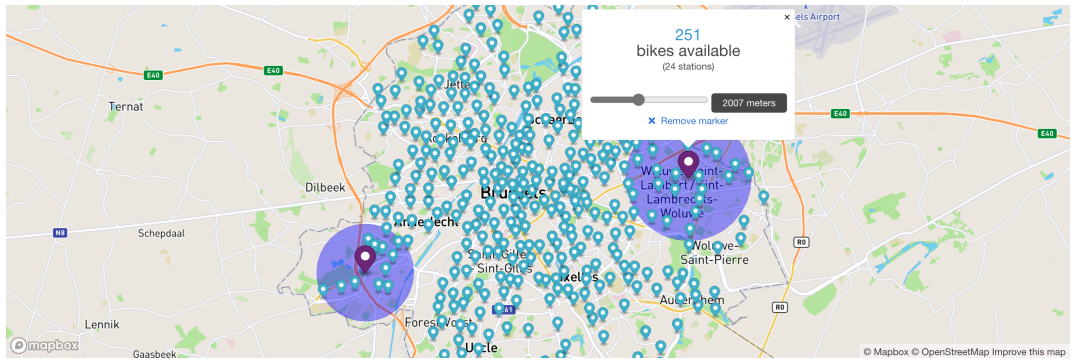

■ **Figure 9** Screenshot of the *Villo!* application. The popup reads "251 bikes available (24 stations) 2007 meters".

## 9 Case Study: Villo!

To evaluate the viability of flocks in a real prosumer application, we implemented a variation of Whereabikes (Section 3) based on real-world data from Villo!, the public bike-sharing programme of Brussels, Belgium.[2] The Villo! API only provides data about its ∼352 stations where bikes are picked up and dropped off, rather than the bikes themselves. Thus, we will count how many bikes are available for pickup at any given time. A screenshot of the application running in a normal web browser is shown in Figure 9, where each small blue marker represents a bike station.

Villo! bike stations are simulated. For each station, an actor replays the events of bikes being taken and returned, which we recorded for an entire day. We evenly distributed these actors over a computer cluster of 160 Raspberry Pis (version 3, model B), a small single-board computer (a photo can be found in Appendix C Figure 11). At the time of writing, 147 of Raspberry Pis are operational for our experiments, each of which simulated 2 or 3 bike stations.

To simulate the application with a larger load, we also ran it on a real cluster that consists of 10 computers with an Intel® Xeon® E3-1240 v5 CPU and 32GB of RAM. Here, we simulated 3000 bike stations, which is approximately the largest bike sharing program in the world (2965 stations in Hangzhou, China [45]). We reused Villo! data by simulating the same stations multiple times but with random perturbations in the data. At this point interaction with the web browser became very slow due to the large number of stations (see Appendix D, Figure 12).

**Lessons Learned**    We faced some practical network-related challenges when using flocks on the clusters.

**Simultaneous connections:** Whenever a new peer joins the network, all other peers with the same flocks simultaneously try to open a connection to the newly discovered peer. To prevent overwhelming the new peer (causing it to crash), connections have to be spread out over time, e.g., by introducing random delays.

---

[2] Villo! — https://www.villo.be/en/home, accessed on 2022-01-11.

**Bidirectional discovery:** When a new peer is discovered via a flock, situations can arise where 2 peers simultaneously open a socket connection to each other. Since WebRTC sockets are bidirectional, a sufficient solution is to have the peers deterministically decide on which of 2 sockets to use (e.g., using a peer's id), and which to close.

## 10 Related Work

In Section 2 we demonstrated the problems that arise when performing acquaintance maintenance and discovery by using RxJS [31, 36] to represent stream-based reactive programming, and REScala [26, 37] to represent behaviour-based reactive programming. While we consider them to be representative for the state of the art, we briefly discuss other related work.

### 10.1 Acquaintance Maintenance

Besides RxJS, other stream-based reactive programming frameworks that we considered include Akka Streams [35], FS2 [14], Project Reactor [29], and Streamz [41]. We found that there is a lot of "cross-pollination" between these frameworks, leading to a similar set of built-in stream operators such as flatMap which we used in Section 2. As such, the problems described using RxJS often translate to other frameworks. One exception is Akka Streams, which also offers powerful operators such as StatefulMap-Concat [19] and "dynamic hubs" [18]. However, the original problem remains that they can be used to craft *custom* solutions for acquaintance maintenance, which is difficult and error-prone.

Most behaviour-based reactive programming languages and libraries offer no mechanisms to perform acquaintance maintenance. Efficient acquaintance maintenance is offered by Scala.React [23] and Hokko [32], but they do not support distributed reactive programming. Support for inefficient acquaintance maintenance can also be found in REScala [26, 37], ScalaLoci [44] and AmbientTalk/R [7].

### 10.2 Acquaintance Discovery

Some related work on reactive programming offers intensional acquaintance discovery.

AmbientTalk/R [7] offers *ambient behaviours* which we showed in Section 2, which have the same purpose as flocks. The added value of flocks is their integration within the reactive program via snapshots and patches.

ScalaLoci [44] is a multi-tier reactive programming language where the various tiers in the distributed application are connected via *ties*. Using our terminology, one acquaintance can be *tied* to multiple other acquaintances.

Potato [42] is a stream-based reactive programming library in Elixir for programming Internet of Things devices, which supports intensional acquaintance discovery. The library provides a global stream of acquaintance "join" and "leave" events, which the programmer can use to manually maintain their own collections of acquaintances.

Akka Streams can leverage the underlying Akka actor library to discover acquaintances. Akka's `Receptionist` can be used to intensionally discover other actors on the network [17], which can be retrieved as a collection of acquaintances.

Other mechanisms exist in non-reactive languages and frameworks, such as AmbientTalk's *volatile sets* [16] and *flocks* [5] for mobile ad-hoc networks. Interaction with these abstractions is based on traditional callbacks, which reactive programming is designed to avoid (due to *callback hell* and *inversion of control* [2, 10]).

## 11 Conclusion

In this paper we analysed the problems of acquaintance management that occur when building reactive prosumer applications, due to 2 levels of reactivity that need to be combined in a correct and efficient way: application-level and topology-level reactivity. Whereas existing reactive programming languages and frameworks offer efficient application-level reactivity, prosumer applications require both to be efficient.

Our solution to acquaintance discovery is the flock. The key to tying flocks and functional reactive programming together is the mechanism of reactor deployments, which we used as a foundation to build an efficient topology-reactive operation called `deploy-*`. The combination of flocks and `deploy-*` resulted in computations that can correctly and efficiently update for both application-level and topology-level reactivity.

**Acknowledgements** We would like to thank James Noble and the anonymous reviewers for their comments and feedback on this paper. Sam Van den Vonder is funded by the Research Foundation – Flanders (FWO) grant No. 1S95318N. Thierry Renaux is funded by the Flanders Innovation & Entrepreneurship (VLAIO) "Cybersecurity Initiative Flanders" program.

■ **Listing 13**  A buggy version of topology-level reactivity in RxJS, calculating an average of all sensors.

```
1  const average$ = sensorDiscoveryService.sensors$.pipe(
2    switchMap((allSensors) =>
3      from(allSensors).pipe(
4        flatMap((sensor) => sensor.value$.pipe(map((val) => [sensor.id, val]))),
5        scan((avgTracker, [id, val]) => avgTracker.update(id, val), new AverageTracker()),
6        map((avgTracker) => avgTracker.getAverage()))));
```

## A  Topology-Level Reactivity in RxJS: A Semantic Bug

An earlier version of this paper presented a different solution to the problem of averaging thermometer measurements used in Section 2. The solution contained a semantic bug. We believe it is realistic that other developers will make same mistake as well, possibly leading to a solution that contains such a bug. Arguably, we think that the (at least) two different ways to think about the problem may actually be part of the problem itself, i.e., part of the problem of accidental complexity when using topology-level reactivity. In this section we explain our buggy implementation, and at the end explain why it is buggy.

We explain the main idea using the same average computation from Section 2 that computes the average value of all sensors (e.g. thermometers) connected to a network. Listing 13 implements this computation via an RxJS stream called average$ ($ is a naming convention for streams). To communicate the continuously appearing and disappearing sensors to the reactive program we used a stream called sensorDiscoveryService.sensors$ (not defined here) from which average$ is derived. This stream propagates an updated Set of all sensors every time a sensor appears or disappears. In RxJS, the average$ stream is defined by "piping" the values from sensors$ through a sequence of RxJS stream operators. The operators in the example work as follows:

**Line 2, switchMap:** The lambda given as its argument will generate a new stream whenever a new set of sensors is propagated. The values produced by switchMap will echo the values of the most recently generated stream, i.e., the average of all sensor values. The body of the lambda generates a new stream via the from operator (line 3) which takes a collection of sensors as input, and returns a stream that emits the elements in the collection one by one.

**Line 4, flatMap:** The lambda given as argument to flatMap generates a stream. flatMap remembers all streams it has generated and echoes their values on a first-in first-out basis. We use it to transform a stream of sensors to a stream of tuples [id, val] where id is the unique identifier of the sensor and val is its latest value. Whenever the value$ stream of a sensor is updated by said sensor, a new [id, val] tuple is propagated by flatMap.

**Line 5, scan:** Most of the actual application logic is tackled by the scan operator that essentially implements a fold operation for streams where the accumulator is emitted every time the input changes. The accumulator is a purely functional

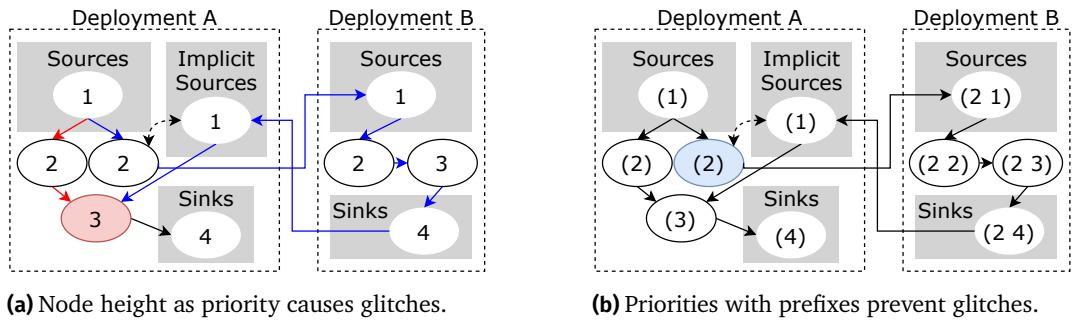

**(a)** Node height as priority causes glitches.

**(b)** Priorities with prefixes prevent glitches.

■ **Figure 10** Side-by-side examples of node priorities that can cause glitches (left), and our solution which does not (right).

`AverageTracker` (implemented elsewhere) that tracks the latest value produced by each sensor. The subsequent `map` (line 6) extracts the current average.

This code incorrectly implements the desired computation. The bug is caused by the propagation of a `Set` of sensors – a solution that is conceptually similar to the solution in behaviour-based reactive programming languages (see Section 2.2.2). This is because an entirely new `Set` of sensors is propagated every time a sensor is added or removed by the discovery service, and the entire result is recomputed from scratch. This means that, internally, all dependencies to the "old" sensors are removed, and new dependencies are established to the "new" sensors (despite their overlap). Because the sensors are connected via a network, this swapping of dependencies would currently cause unnecessary network traffic and delays. Worse, every time the set of sensors changes, the final output stream emits all intermediate values until all sensors are known (again) to `flatMap`, which is semantically incorrect.

## B Glitch Prevention for Deployments

Reactive programming languages usually use a smart propagation algorithm that prevents *glitches*, which are temporary inconsistencies in the state of the reactive program that result in incorrect and redundant computations [9]. Reactive programming languages such as REScala [37] and Flapjax [25] use a propagation algorithm based on FrTime [9], where glitches are avoided by assigning a height to every node in the DAG and using those heights to schedule nodes for re-execution in a priority queue. However, their propagation algorithms no longer prevent glitches in the context of Stella with reusable reactor behaviours and deployments. While the implementation of Stella is outside of the scope of this paper, we have discovered that a propagation algorithm in the same style as FrTime that can still be used by changing the way node priorities are calculated.

Consider the DAG of an arbitrary reactive program depicted in Figure 10a, where every node is labelled with its statically computed height in the DAG of their reactor behaviour (note that the dashed-line boxes are deployments of two different reactor behaviours). The DAG of deployment A contains a `deploy` expression that results in

■ **Listing 14**  TypeScript procedure to determine which priority list has the highest priority.

```typescript
function hasHigherPriority(priorities1: number[], priorities2: number[]): boolean {
    for (let i = 0; i < Math.min(priorities1.length, priorities2.length); i++) {
        if (priorities1[i] < priorities2[i])
            return true;
    }
    return priorities1.length < priorities2.length;
}
```

deployment B (the first "2" node on the blue path). Similar to a qualification and `deploy-*`, a `deploy` expression is also split up into a node that manages the created deployment, and an implicit source node that is responsible for receiving the values from the created deployment. When nodes are scheduled according to the original propagation algorithm, a glitch occurs in the node highlighted in red. Concretely, its computation is executed two times instead of only once. The first computation is redundant and should be avoided. As a visual aid, we highlighted the two paths of execution in red and blue that both cause an update of the red node at different moments in the propagation cycle.

Nodes with the same height (priority) may be executed in an arbitrary order, and a lower number denotes a higher priority. The problem occurs when the source node of deployment A has a new value, and gradually all dependents on the red and blue paths are scheduled. Since the red node has a priority of 3, it is computed once after all nodes of priority 2 have been computed. However, via the blue path it indirectly depends on a sink node with the lower priority 4. Whenever this sink node changes, it will schedule its dependents for recomputation, and eventually the red node is recomputed once again.

Our solution is to ensure that there is a globally correct ordering between nodes across the different deployments within the same reactor. Instead of using a single number (height) to represent priority, priorities now consist of a list of heights which includes the priority of the predecessing deployment(s). Consider the diagram in Figure 10b, which is the same diagram as Figure 10a but using our system of assigning priorities. If we assume deployment A is the root deployment of the reactor, then the priorities of its nodes are now represented by singleton lists with the same heights as Figure 10a. Whenever a new deployment is created, such as deployment B which is created by the blue `deploy` node, then the priorities of all nodes in deployment B receive a prefix which is equal to the priority of the node that caused the additional deployment. In this case the prefix is the priority list (2), which is added in front of the pre-computed heights of the nodes in deployment B.

To determine the position in the priority queue, values in the priority lists are compared left-to-right according to the TypeScript procedure in Listing 14. The function `hasHigherPriority` returns whether its first argument `priorities1` has a higher priority than its second argument `priorities2`, in which case it returns true. The values in both priority lists are compared pairwise from left to right. Then, `priorities1` has a higher priority whenever one of its values is smaller. For example, (1 2) has a higher priority

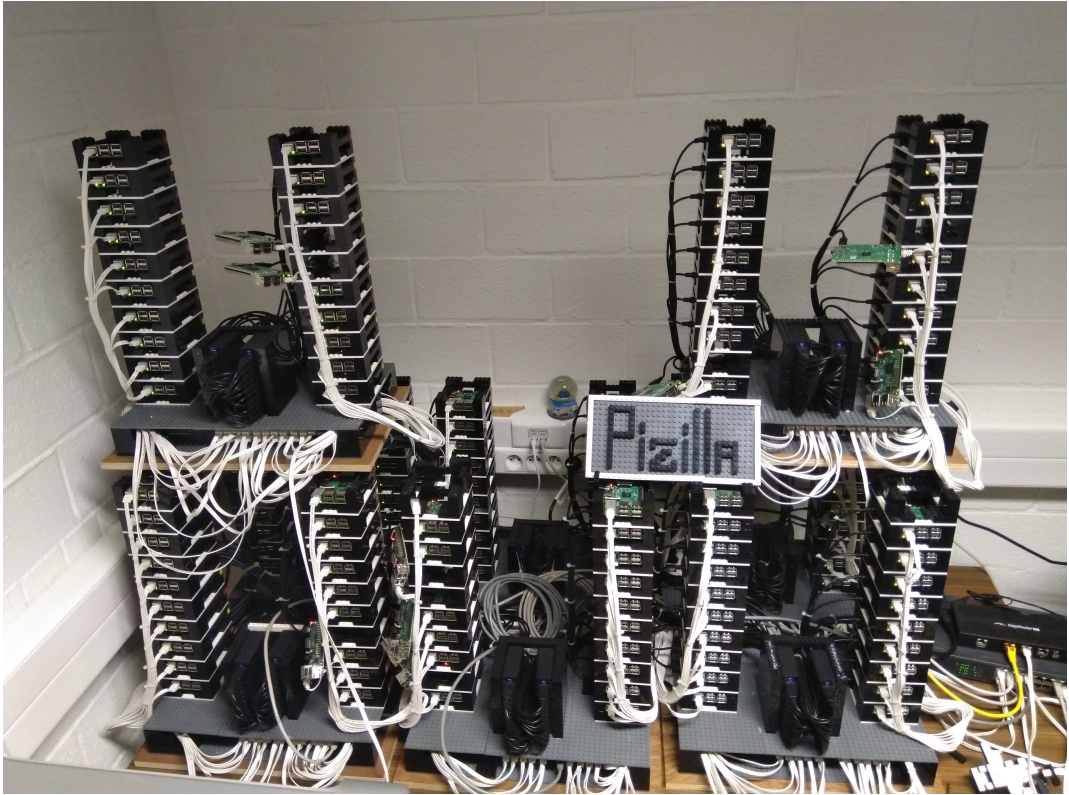

■ **Figure 11**    Photo of the Raspberry Pi cluster. The Raspberry Pis are encased in LEGO®.

than (2 2) because the first priority in the first priority list (1) is smaller than the first priority in the second priority list (2). When the priorities are the same up until the length of the shortest array, then the shortest priority list has the highest priority. For instance, using the priorities from Figure 10b, (2) has priority over (2 1) to (2 4).

## C    Raspberry Pi Cluster (Generating New Acquaintances)

Experiments were run on a cluster of 160 Raspberry Pis (version 3, model B), a small single-board computer. This cluster is usually used for research on distributed systems. A photo can be found in Figure 11.

## D    Intel Xeon Cluster (Acquaintance Management)

Figure 12 presents a screenshot of the Villo! application with 3000 simulated stations. Since we introduced perturbations in the data of the Villo! stations (including their location), the city of Brussels is completely covered in station markers. Due to the large number of map markers, the browser had difficulties both rendering the map and running the Stella interpreter in the background.

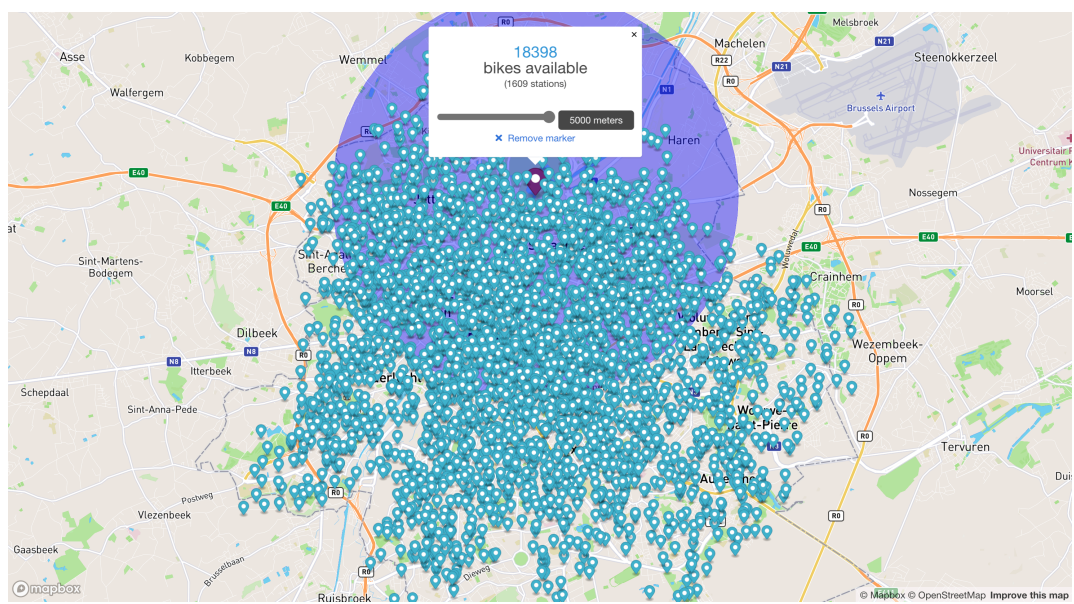

■ **Figure 12** Screenshot of the *Villo!* application with 3000 simulated bike stations. The pop-up reads "18398 bikes available (1609 stations) 5000 meters".

## References

[1]   Feross Aboukhadijeh. *simple-peer: Simple WebRTC video, voice, and data channels*. http://web.archive.org/web/20210216115846/https://github.com/feross/simple-peer. Accessed on 2021-02-16.

[2]   Engineer Bainomugisha, Andoni Lombide Carreton, Tom Van Cutsem, Stijn Mostinckx, and Wolfgang De Meuter. "A survey on reactive programming". In: *ACM Computing Surveys* 45.4 (2013), 52:1–52:34. DOI: 10.1145/2501654.2501666.

[3]   Herman Banken, Erik Meijer, and Georgios Gousios. "Debugging Data Flows in Reactive Programs". In: *Proceedings of the 40th International Conference on Software Engineering, ICSE 2018, Gothenburg, Sweden, May 27 - June 03, 2018*. Edited by Michel Chaudron, Ivica Crnkovic, Marsha Chechik, and Mark Harman. ICSE '18. Gothenburg, Sweden: Association for Computing Machinery, 2018, pages 752–763. ISBN: 978-1-4503-5638-1. DOI: 10.1145/3180155.3180156.

[4]   Edd Barrett, Carl Friedrich Bolz-Tereick, Rebecca Killick, Sarah Mount, and Laurence Tratt. "Virtual machine warmup blows hot and cold". In: *Proc. ACM Program. Lang.* 1.OOPSLA (2017), 52:1–52:27. DOI: 10.1145/3133876.

[5]   Elisa Gonzalez Boix, Andoni Lombide Carreton, Christophe Scholliers, Tom Van Cutsem, Wolfgang De Meuter, and Theo D'Hondt. "Flocks: Enabling Dynamic Group Interactions in Mobile Social Networking Applications". In: *Proceedings of the 2011 ACM Symposium on Applied Computing*. SAC '11. TaiChung, Taiwan: Association for Computing Machinery, 2011, pages 425–432. ISBN: 9781450301138. DOI: 10.1145/1982185.1982277.

[6]   Jan-Ivar Bruaroey, Cullen Jennings, and Henrik Boström. *WebRTC 1.0: Real-Time Communication Between Browsers*. W3C Recommendation. http://web.archive.org/web/20210210094400/https://www.w3.org/TR/2021/REC-webrtc-20210126/, Accessed on 2021-02-10. W3C, Jan. 2021.

[7]   Andoni Lombide Carreton, Stijn Mostinckx, Tom Van Cutsem, and Wolfgang De Meuter. "Loosely-Coupled Distributed Reactive Programming in Mobile Ad Hoc Networks". In: *Objects, Models, Components, Patterns, 48th International Conference, TOOLS 2010, Málaga, Spain, June 28 - July 2, 2010. Proceedings*. Edited by Jan Vitek. Volume 6141. Lecture Notes in Computer Science. Springer, 2010, pages 41–60. DOI: 10.1007/978-3-642-13953-6_3.

[8]   Gregory H. Cooper. "Integrating Dataflow Evaluation into a Practical Higher-Order Call-by-Value Language". PhD thesis. Providence, Rhode Island: Brown University, May 2008.

[9]   Gregory H. Cooper and Shriram Krishnamurthi. "Embedding Dynamic Dataflow in a Call-by-Value Language". In: *Programming Languages and Systems, 15th European Symposium on Programming, ESOP 2006, Held as Part of the Joint European Conferences on Theory and Practice of Software, ETAPS 2006, Vienna, Austria, March 27-28, 2006, Proceedings*. Edited by Peter Sestoft. Volume 3924. Lecture Notes in Computer Science. Springer, 2006, pages 294–308. DOI: 10.1007/11693024_20.

[10]  Jonathan Edwards. "Coherent reaction". In: *Companion to the 24th Annual ACM SIGPLAN Conference on Object-Oriented Programming, Systems, Languages, and Applications, OOPSLA 2009, October 25-29, 2009, Orlando, Florida, USA*. Edited by Shail Arora and Gary T. Leavens. ACM, 2009, pages 925–932. ISBN: 978-1-60558-768-4. DOI: 10.1145/1639950.1640058.

[11]  Conal Elliott and Paul Hudak. "Functional Reactive Animation". In: *Proceedings of the 1997 ACM SIGPLAN International Conference on Functional Programming (ICFP '97), Amsterdam, The Netherlands, June 9-11, 1997*. Edited by Simon L. Peyton Jones, Mads Tofte, and A. Michael Berman. ACM, 1997, pages 263–273. DOI: 10.1145/258948.258973.

[12]  U.S. Census Bureau Geographic Information Systems FAQ. *GIS FAQ Q5.1: Great circle distance between 2 points*. http://web.archive.org/web/20201029002559/http://www.movable-type.co.uk/scripts/gis-faq-5.1.html. Accessed on 2021-02-10.

[13]  Daniel P. Friedman and Mitchell Wand. *Essentials of programming languages*. 3rd edition. MIT Press, 2008. ISBN: 978-0-262-06279-4.

[14]  *FS2: The Official Guide*. http://web.archive.org/web/20200824074626/https://fs2.io/guide.html. Accessed on 2020-08-24.

[15]  Ilya Grigorik. *High Resolution Time Level 2*. W3C Recommendation. Accessed on 2021-03-25. W3C, Nov. 2019. URL: https://web.archive.org/web/20210325154905/https://www.w3.org/TR/2019/REC-hr-time-2-20191121/.

[16]  Dries Harnie, Elisa Gonzalez Boix, Andoni Lombide Carreton, Christophe Scholliers, and Wolfgang De Meuter. "Volatile Sets: Event-driven Collections for Mobile Ad-Hoc Applications". In: *Electronic Communications of the EASST* 43 (2011). DOI: 10.14279/tuj.eceasst.43.585.

[17]  Lightbend Inc. *Actor discovery – Akka Documentation*. Accessed on 2022-01-11. URL: http://web.archive.org/web/20220111091555/https://doc.akka.io/docs/akka/2.5.32/typed/actor-discovery.html.

[18]  Lightbend Inc. *Akka Documentation: Dynamic fan-in and fan-out with MergeHub, BroadcastHub and PartitionHub*. http://web.archive.org/web/20200821094744/https://doc.akka.io/docs/akka/current/stream/stream-dynamic.html. Accessed on 2020-08-21.

[19]  Lightbend Inc. *Akka Documentation: statefulMapConcat*. http://web.archive.org/web/20210804174155/https://doc.akka.io/docs/akka/current/stream/operators/Source-or-Flow/statefulMapConcat.html. Accessed on 2021-12-10.

[20]  Optimal Strategix Group Inc. *The customer is dead: long live the Prosumer*. http://web.archive.org/web/20210112084517/https://www.osganalytics.com/the-customer-is-dead-long-live-the-prosumer/. Accessed on 2021-01-12.

[21]  Rick Kawamura. *How Hard is it to Build an IoT Electric Scooter Fleet like Bird or Lime?* http://web.archive.org/web/20210119133924/https://www.soracom.io/blog/how-hard-is-it-to-build-an-electric-scooter-fleet-like-bird-or-lime/. Accessed on 2021-01-19. Sept. 2019.

[22]  Joeri De Koster, Tom Van Cutsem, and Wolfgang De Meuter. "43 years of actors: a taxonomy of actor models and their key properties". In: *Proceedings of the 6th International Workshop on Programming Based on Actors, Agents, and Decentralized Control, AGERE 2016, Amsterdam, The Netherlands, October 30, 2016*. Edited by Sylvan Clebsch, Travis Desell, Philipp Haller, and Alessandro Ricci. ACM, 2016, pages 31–40. DOI: 10.1145/3001886.3001890.

[23]  Ingo Maier and Martin Odersky. "Higher-Order Reactive Programming with Incremental Lists". In: *ECOOP 2013 - Object-Oriented Programming - 27th European Conference, Montpellier, France, July 1-5, 2013. Proceedings*. Edited by Giuseppe Castagna. Volume 7920. Lecture Notes in Computer Science. Springer, 2013, pages 707–731. DOI: 10.1007/978-3-642-39038-8_29.

[24]  Alessandro Margara and Guido Salvaneschi. "On the Semantics of Distributed Reactive Programming: The Cost of Consistency". In: *IEEE Trans. Software Eng.* 44.7 (2018), pages 689–711. DOI: 10.1109/TSE.2018.2833109.

[25]  Leo A. Meyerovich, Arjun Guha, Jacob P. Baskin, Gregory H. Cooper, Michael Greenberg, Aleks Bromfield, and Shriram Krishnamurthi. "Flapjax: a programming language for Ajax applications". In: *Proceedings of the 24th Annual ACM SIGPLAN Conference on Object-Oriented Programming, Systems, Languages, and Applications, OOPSLA 2009, October 25-29, 2009, Orlando, Florida, USA*. Edited by Shail Arora and Gary T. Leavens. ACM, 2009, pages 1–20. DOI: 10.1145/1640089.1640091.

[26]   Ragnar Mogk, Lars Baumgärtner, Guido Salvaneschi, Bernd Freisleben, and
       Mira Mezini. "Fault-tolerant Distributed Reactive Programming". In: *32nd Eu-
       ropean Conference on Object-Oriented Programming, ECOOP 2018, July 16-21,
       2018, Amsterdam, The Netherlands*. Edited by Todd D. Millstein. Volume 109.
       LIPIcs Leibniz International Proceedings in Informatics. Schloss Dagstuhl -
       Leibniz-Zentrum für Informatik, 2018, 1:1–1:26. DOI: 10.4230/LIPIcs.ECOOP.2018
       .1.

[27]   Florian Myter, Christophe Scholliers, and Wolfgang De Meuter. "Distributed
       Reactive Programming for Reactive Distributed Systems". In: *The Art, Science,
       and Engineering of Programming* 3.3 (2019), page 5. DOI: 10.22152/programming-
       journal.org/2019/3/5.

[28]   José Proença and Carlos Baquero. "Quality-Aware Reactive Programming for
       the Internet of Things". In: *Fundamentals of Software Engineering - 7th Interna-
       tional Conference, FSEN 2017, Tehran, Iran, April 26-28, 2017, Revised Selected
       Papers*. Edited by Mehdi Dastani and Marjan Sirjani. Volume 10522. Lecture
       Notes in Computer Science. Springer, 2017, pages 180–195. DOI: 10.1007/978-3
       -319-68972-2_12.

[29]   *Project Reactor: Create Efficient Reactive Systems*. http://web.archive.org/web/20
       200817051359/https://projectreactor.io/. Accessed on 2020-09-08.

[30]   Belga / N. Quintelier. *Villo lanceert 1.800 elektrische stadsfietsen in Brussel*.
       http://web.archive.org/web/20210119143206/https://nl.metrotime.be/2019/11/30
       /must-read/villo-lanceert-1-800-elektrische-stadsfietsen-in-brussel/. Accessed
       on 2021-01-19. Nov. 2019.

[31]   *ReactiveX: An API for asynchronous programming with observable streams*. http:
       //web.archive.org/web/20191009085652/http://reactivex.io/. Accessed on
       2019-10-09.

[32]   Bob Reynders and Dominique Devriese. "Efficient Functional Reactive Pro-
       gramming Through Incremental Behaviors". In: *Programming Languages and
       Systems - 15th Asian Symposium, APLAS 2017, Suzhou, China, November 27-29,
       2017, Proceedings*. Edited by Bor-Yuh Evan Chang. Volume 10695. Lecture Notes
       in Computer Science. Springer, 2017, pages 321–338. DOI: 10.1007/978-3-319-712
       37-6_16.

[33]   Bob Reynders, Dominique Devriese, and Frank Piessens. "Multi-Tier Functional
       Reactive Programming for the Web". In: *Onward! 2014, Proceedings of the 2014
       ACM International Symposium on New Ideas, New Paradigms, and Reflections on
       Programming & Software, part of SPLASH '14, Portland, OR, USA, October 20-24,
       2014*. Edited by Andrew P. Black, Shriram Krishnamurthi, Bernd Bruegge, and
       Joseph N. Ruskiewicz. ACM, 2014, pages 55–68. DOI: 10.1145/2661136.2661140.

[34]   George Ritzer and Nathan Jurgenson. "Production, Consumption, Prosumption:
       The nature of capitalism in the age of the digital 'prosumer'". In: *Journal of
       Consumer Culture* 10.1 (2010), pages 13–36. DOI: 10.1177/1469540509354673.

[35]   Raymond Roestenburg, Rob Bakker, and Rob Williams. "Akka in action". In:
       1st edition. Manning Publications Co., 2016. Chapter 13. ISBN: 978-1-61729-101-2.

[36]  *RxJS: Reactive Extensions For JavaScript*. http://web.archive.org/web/202008200
80335/https://github.com/ReactiveX/rxjs. Accessed on 2020-08-20.

[37]  Guido Salvaneschi, Gerold Hintz, and Mira Mezini. "REScala: bridging be-
tween object-oriented and functional style in reactive applications". In: *13th
International Conference on Modularity, MODULARITY '14, Lugano, Switzerland,
April 22-26, 2014*. Edited by Walter Binder, Erik Ernst, Achille Peternier, and
Robert Hirschfeld. ACM, 2014, pages 25–36. DOI: 10.1145/2577080.2577083.

[38]  Guido Salvaneschi and Mira Mezini. "Debugging for Reactive Programming".
In: *Proceedings of the 38th International Conference on Software Engineering*.
Edited by Laura K. Dillon, Willem Visser, and Laurie A. Williams. ICSE '16.
Austin, Texas: Association for Computing Machinery, 2016, pages 796–807.
ISBN: 9781450339001. DOI: 10.1145/2884781.2884815.

[39]  Guido Salvaneschi, Sebastian Proksch, Sven Amann, Sarah Nadi, and Mira
Mezini. "On the Positive Effect of Reactive Programming on Software Com-
prehension: An Empirical Study". In: *IEEE Trans. Software Eng.* 43.12 (2017),
pages 1125–1143. DOI: 10.1109/TSE.2017.2655524.

[40]  Kazuhiro Shibanai and Takuo Watanabe. "Distributed functional reactive pro-
gramming on actor-based runtime". In: *Proceedings of the 8th ACM SIGPLAN
International Workshop on Programming Based on Actors, Agents, and Decen-
tralized Control, AGERE!@SPLASH 2018, Boston, MA, USA, November 5, 2018*.
Edited by Joeri De Koster, Federico Bergenti, and Juliana Franco. ACM, 2018,
pages 13–22. DOI: 10.1145/3281366.3281370.

[41]  *Streamz: Streamz Documentation*. http://web.archive.org/web/20200908160021
/https://streamz.readthedocs.io/en/latest/. Accessed on 2020-09-08.

[42]  Christophe De Troyer, Jens Nicolay, and Wolfgang De Meuter. "Building IoT
Systems Using Distributed First-Class Reactive Programming". In: *2018 IEEE
International Conference on Cloud Computing Technology and Science, CloudCom
2018, Nicosia, Cyprus, December 10-13, 2018*. IEEE Computer Society, 2018,
pages 185–192. DOI: 10.1109/CloudCom2018.2018.00045.

[43]  Sam Van den Vonder, Thierry Renaux, Bjarno Oeyen, Joeri De Koster, and
Wolfgang De Meuter. "Tackling the Awkward Squad for Reactive Programming:
The Actor-Reactor Model". In: *34th European Conference on Object-Oriented
Programming, ECOOP 2020, November 15-17, 2020, Berlin, Germany (Virtual
Conference)*. Edited by Robert Hirschfeld and Tobias Pape. Volume 166. LIPIcs.
Schloss Dagstuhl - Leibniz-Zentrum für Informatik, 2020, 19:1–19:29. DOI: 10.42
30/LIPIcs.ECOOP.2020.19.

[44]  Pascal Weisenburger, Mirko Köhler, and Guido Salvaneschi. "Distributed System
Development with ScalaLoci". In: *Proc. ACM Program. Lang.* 2.OOPSLA (Feb.
2018). DOI: 10.1145/3276499.

[45]  Wikipedia. *List of bicycle-sharing systems — Wikipedia, The Free Encyclopedia*.
http://web.archive.org/web/20210402144924/https://en.wikipedia.org/wiki/List_
of_bicycle-sharing_systems. Accessed on 2021-04-02. 2021.

## About the authors

**Sam Van den Vonder** is a Ph D student at the Software Languages Lab of the Vrije Universiteit Brussel, and the designer of the actor-based reactive programming language called Stella. His main research area is (distributed) reactive programming, and more concretely studying the semantics of the interactions between reactive and non-reactive parts of programs. Contact him at Sam.Van.den.Vonder@vub.be.
**ORCID:** 0000-0002-9241-1098

**Thierry Renaux** is a postdoctoral researcher at the Software Languages Lab of the Vrije Universiteit Brussel. His current research interests include security by/for reactive systems. Contact him at Thierry.Renaux@vub.be.
**ORCID:** 0000-0002-9301-2187

**Wolfgang De Meuter** is a professor at the Vrije Universiteit Brussel, is the head of the Software Languages Lab, and is specialised in programming languages and programming tools. His current research is situated in the field of distributed programming, concurrent programming, reactive programming, and big data processing. Contact him at Wolfgang.De.Meuter@vub.be.
**ORCID:** 0000-0002-5229-5627